\documentclass[
aps,
prc,
superscriptaddress,
floatfix,
nofootinbib,
twocolumn
]{revtex4-2}

\usepackage{amsmath,amssymb,amsfonts,physics,graphicx,float}
\graphicspath{{pic/}}
\usepackage[setpagesize=false]{hyperref}
\allowdisplaybreaks[4]

\usepackage[labelformat=simple]{subcaption} 

\usepackage{ulem}
\usepackage{color}
\definecolor{ar}{rgb}{1.0, 0.01, 0.24}
\definecolor{al}{rgb}{0.82, 0.1, 0.26}
\definecolor{ev}{rgb}{0.56, 0.0, 1.0}

\begin{document}

\title{
Reconciling the HESS J1731-347 constraints with \\
Parity doublet model
}

\author{Bikai Gao}
\email{gaobikai@hken.phys.nagoya-u.ac.jp}
\affiliation{Department of Physics, Nagoya University, Nagoya 464-8602, Japan}

\author{Yan Yan}
\email{2919ywhhxh@163.com}
\affiliation{School of Microelectronics and Control Engineering, Changzhou University, Jiangsu 213164, China}

\author{Masayasu Harada}
\email{harada@hken.phys.nagoya-u.ac.jp}
\affiliation{Kobayashi-Maskawa Institute for the Origin of Particles and the Universe, Nagoya University, Nagoya, 464-8602, Japan}
\affiliation{Department of Physics, Nagoya University, Nagoya 464-8602, Japan}
\affiliation{Advanced Science Research Center, Japan Atomic Energy Agency, Tokai 319-1195, Japan}

\date{\today}

\begin{abstract}
The recent discovery of a central compact object (CCO) within the supernova remnant HESS J1731-347, characterized by a mass of approximately $0.77^{+0.20}_{-0.17} M_{\odot}$ and a radius of about $10.4^{+0.86}_{-0.78}$ km, has opened up a new window for the study of compact objects. This CCO is particularly intriguing because it is the lightest and smallest compact object ever observed, raising questions and challenging the existing theories.
To account for this light compact star, a mean-field model within the framework of parity doublet structure is applied to describe the hadron matter. Inside the model, part of the nucleon mass is associated with the chiral symmetry breaking while the other part  is  
from the chiral invariant mass $m_0$ which is insensitive to the temperature/density. The value of $m_0$ affects the nuclear equation of state for uniform nuclear matter at low density and exhibits strong correlations with the radii of neutron stars. 
 We point out  
that HESS J1731-347 can be explained as  
the lightest neutron star  for $m_0 \simeq 850 \,$MeV.

\end{abstract}

\maketitle


\section{Introduction}
Neutron star (NS) is one of the most compact objects in the universe with a mass of $1$-$2M_{\odot}$ and a radius of $\sim 10$\,km. The NSs with extreme conditions provide us unique natural laboratory for investigating the phases of cold, dense matter, including the possibility of exotic states such as hyperons and even quarks appearing within these astrophysical objects. Understanding the properties of NSs requires the information about its equation of state (EOS) which characterizes how pressure $P$ varies as a function of energy density $\epsilon$. This EOS cannot be directly predicted by the quantum chromodynamics (QCD) and also the lattice QCD simulations due to the sign problem.  Thanks to the advancements of recent multi-messenger astronomy on different sources, especially those made by gravitational wave laser interferometers from the LIGO-VIRGO\cite{PhysRevLett.119.161101,LIGOScientific:2017ync,LIGOScientific:2018cki} and X-ray emissions observations conducted by the Neutron Star Interior Composition Explorer (NICER), we made remarkable improvements to constrain the EOS of cold, dense and strongly interacting nuclear matter. For instance, the NS merger event GW170817 provided insights into the mass and radius of NSs, with an estimation of approximately 1.4$M_{\odot}$ and a radius of $R = 11.9 ^{+1.4}_{-1.4}$ km. This observation suggested that the EOS should be relatively soft for uniform nuclear matter existing in the low-density region. Additionally, NICER has played a crucial role in advancing our understandings of NSs. The analyses\cite{Miller:2021qha, Riley:2021pdl} have focused on NSs with masses around $1.4M_{\odot}$ and $\sim 2.1 M_{\odot}$. Interestingly, the results indicated that the radii of these NSs are rather similar for different masses, with a radius of approximately $12.45 \pm 0.65$ kilometers for a 1.4 $M_{\odot}$ NS and $12.35 \pm 0.75$ kilometers for a 2.08 $M_{\odot}$ NS. These findings suggest that the EOS stiffens rapidly, meaning that the pressure increases quickly as a function of energy density, as one moves from low baryon density ($\lesssim 2n_{0}$; $n_0$:  nuclear saturation density ) to high density ($4$-$7n_0$). This stiffening of the EOS is necessary to support the existence of massive NSs, such as those with masses around 2$M_{\odot}$.


The recent report on the central compact object (CCO) HESS J1731-347\cite{2022A} with an estimated mass and radius of the object are $M = 0.77^{+0.20}_{-0.17} M_{\odot}$ and $R = 10.4^{+0.86}_{-0.78}$ km, have raised many questions and put more constraints into the EOS. This measurements suggest that this CCO may correspond to a neutron star with an even softer equation of state in the low-density region than previously observed. Some studies considered the possibility that HESS J1731-347 may be a quark star\cite{Chu:2023rty,Oikonomou:2023otn,Restrepo:2022wqn,Yang:2023haz,Rather:2023tly}, an exotic theoretical object composed of deconfined quarks rather than the usual hadronic matter suggested in neutron stars.

In this research, we will explore the possibility that HESS J1731-347 may be the neutron star within the framework of a quark-hadron crossover model constructed in \cite{Baym:2017whm,Minamikawa:2020jfj,Minamikawa:2021fln,Gao:2022klm}, in which a unified EOS is constructed by interpolating the hadronic EOS from a hadronic model based on the parity doublet structure\cite{PhysRevD.39.2805,10.1143/PTP.106.873}.
 and the quark EOS from an NJL-type quark model.

Hadronic models based on the parity doublet structure, which we call parity doublet models(PDMs), offer a unique perspective on the structure of hadrons by considering the existence of chiral invariant mass, denoted by $m_0$, in addition to the conventional chiral variant mass 
generated by the spontaneous chiral symmetry breaking.  
The existence of the chiral invariant mass is consistent with the lattice QCD simulation done at non-zero temperature  \cite{Aarts:2015mma,Aarts:2017rrl,Aarts:2018glk}.
The framework of PDMs has been widely used to study the hadron structure\cite{Nishihara:2015fka,Chen:2009sf,Chen:2010ba,Chen:2011rh,Minamikawa:2023ypn,Gao:2024mew} and  construct the EOS for nuclear /NS matter\cite{HATSUDA198911,PhysRevC.75.055202,PhysRevC.77.025803,PhysRevC.82.035204,PhysRevD.84.034011,GALLAS201113,PhysRevC.84.045208,PhysRevD.85.054022,PhysRevC.87.015804,PhysRevD.88.105019,PhysRevC.96.025205,PhysRevC.97.045203,PhysRevC.97.065202,refId01,universe5080180,Motohiro,PhysRevC.100.025205,PhysRevC.103.045205,Mukherjee:2017jzi,doi:10.7566/JPSCP.26.024001,Gao:2022klm,Marczenko:2022hyt,Minamikawa:2023eky,Baym:2017whm,Minamikawa:2020jfj,Minamikawa:2021fln,Gao:2022klm,Minamikawa:2023eky,Marczenko:2019trv,Kong:2023nue}. We note that the constructed EOS is softer for larger chiral invariant mass, and the resultant EOSs are combined with the EOS constructed from an NJL-type quark model by assuming  quark-hadron crossover,  which allows for a smooth transition from hadronic matter to quark matter\cite{Baym:2017whm,Minamikawa:2020jfj,Minamikawa:2021fln,Gao:2022klm,Minamikawa:2023eky,Marczenko:2019trv}. This hybrid approach, where the PDM EOS is employed up to densities around $2$-$3n_0$ 
and interpolate with the quark EOS at $\geq 5 n_0$ via polynomial interpolation to obtain the unified EOS. In this case, the unified EOS can be constructed with soft EOS in the low density part and sufficiently stiff EOS in the high density part to support the $2M_{\odot}$ constraint.

 In this work, we consider a hadronic EOS constructed from a  PDM in the low density region and interpolate with quark EOS using an  NJL-type quark model in the high density region. Inside the PDM, we included the $\rho^{2}\omega^{2}$ interaction term with $\lambda_{\omega\rho}$ to be its coupling constant, which is assumed to make the EOS softer. By adjusting the two parameters $\lambda_{\omega\rho}$ and $m_0$, we can adjust the stiffness of EOS in the hadronic model.
 The  
 constructed unified EOS is shown to satisfty  
 the constraints from HESS J1731-347, 
 makes it possible to be the lightest neutron star ever observed.

This paper is organized as follows. In Sec.~\ref{sec-eos}, we explain the formulation of
present model. The main results of the analysis of properties of NS are shown in Sec.~\ref{sec-NS}. Finally, we show the summary and discussions in  Sec.~\ref{summary}.



\section{EQUATION OF STATE }\label{sec-eos}
In this section, we briefly review how to construct neutron star matter EOS from a 
PDM in the low-density region, and from  a  NJL-type quark model in the high-density region. 

\subsection{NUCLEAR MATTER EOS}
\label{sec:PDM matter}

 In Ref.~\cite{Gao:2022klm}, a hadronic parity doublet model (PDM) is constructed to describe the NS properties in the low density region ($\leq 2 n_0$). 
The model includes the effects of strange quark chiral condensate  through the KMT-type interaction in the mesonic sector. 
The density dependence of the strange quark chiral condensate $\langle \bar{s}s \rangle$  is calculated and the results show the impacts of strange quark chiral condensate is very limited in the low density region. Then, in the current study, we neglect the effect of strange quark in the low density domain. In addition, we ignore the influence of the isovector scalar meson $a_0(980)$ in the current model, which believed to appear in asymmetric matter like neutron stars. As investigated in Ref.~\cite{Kong:2023nue}, the effect of the $a_0(980)$ has a negligible impact on the properties of neutron stars. Specifically, the inclusion of the $a_0(980)$ only results in a slight increase in the radius by less than a kilometer. We would like also to note that, in these analyses, a term of vector meson mixing, i.e. $\omega^2\rho^2$ term, is introduced to make the slope parameter to be consistent with the recent constraint shown in Ref.~\cite{universe7060182}. In the present analysis, we also include the mixing contribution.

The thermodynamic potential is obtained as \cite{Motohiro,PhysRevC.103.045205}
\begin{equation}
\begin{aligned}
\Omega_{\mathrm{PDM}}= & V\left(\sigma\right)-V\left(\sigma_0\right)-\frac{1}{2} m_\omega^2 \omega^2-\frac{1}{2} m_\rho^2 \rho^2 \\
& -\lambda_{\omega \rho}\left(g_\omega \omega\right)^2\left(g_\rho \rho\right)^2 \\
& -2 \sum_{i=+,-} \sum_{\alpha=p, n} \int^{k_f} \frac{\mathrm{d}^3 \mathbf{p}}{(2 \pi)^3}\left(\mu_\alpha^*-E_{\mathrm{p}}^i\right)\ ,
\end{aligned}
\label{Omega PDM}
\end{equation}
where $i = +, -$ denote 
the parity of nucleons and $E_{{\bf p}}^{i} = \sqrt{{\bf p}^{2} + m_{i}^{2}}$ is the energy of nucleons with mass $m_{i}$ and momentum ${\bf p}$.
 In Eq~(\ref{Omega PDM}), the  
potential $V(\sigma)$ is given  by
\begin{align}
V(\sigma) = -\frac{1}{2}\bar{\mu}^{2}\sigma^{2} + \frac{1}{4}\lambda_4 \sigma^4 -\frac{1}{6}\lambda_6\sigma^6 - m_{\pi}^{2} f_{\pi}\sigma\ ,
\end{align}
and $\sigma_0$ is the mean field at vacuum.

We note that the sign of $\lambda$ is restricted to be positive due to the stability of the vacuum at zero density\cite{Kong:2023nue}. The total thermodynamic potential for the NS is obtained by including the effects of leptons as
\begin{align}
\Omega_{{\rm H}} = \Omega_{{\rm PDM}} + \sum_{l = e, \mu}\Omega_l \ , 
\end{align}
where $\Omega_{l}(l=e,\mu)$ are the thermodynamic potentials for leptons  given by 
\begin{equation}
\Omega_{l}=-2 \int^{k_{F}} \frac{d^{3} \mathbf{p}}{(2 \pi)^{3}}\left(\mu_{l}-E_{\mathbf{p}}^{l}\right).
\end{equation}
The mean fields here are determined by following stationary conditions:
\begin{equation}
0=\frac{\partial \Omega_{\mathrm{H}}}{\partial \sigma}, \quad 0=\frac{\partial \Omega_{\mathrm{H}}}{\partial \omega}, \quad 0=\frac{\partial \Omega_{\mathrm{H}}}{\partial \rho}.
\end{equation}
We also need to consider the $\beta$ equilibrium and the charge neutrality conditions,
\begin{align}
\mu_{e}=\mu_{\mu}=-\mu_{Q} ,\\
\frac{\partial \Omega_{\mathrm{H}}}{\partial \mu_{Q}}=n_{p}-n_{l}=0 \,,
\end{align}
 where $\mu_Q$ is the charge chemical potential. 
We then have the pressure in hadronic matter as
\begin{equation}
P_{\mathrm{H}}=-\Omega_{\mathrm{H}}.
\end{equation}

We then determine the parameters in the  PDM
by fitting them to the pion decay constant and hadron masses given in Table.~\ref{input: mass} and  the normal nuclear matter properties summarized in Table.~\ref{saturation} for fixed value of $m_0$. 
\begin{table}[htbp]
\centering
	\caption{  {\small Physical inputs in vacuum in unit of MeV.  }  }\label{input: mass}
	\begin{tabular}{cccccc}
		\hline\hline
		~$m_\pi$~&~ $f_\pi$ ~ &~ $m_\omega$ ~&~ $m_\rho$ ~&~ $m_+$ ~&~ $m_-$\\
		\hline
		~140 ~&~ 92.4 ~&~ 783 ~&~ 776 ~&~ 939 ~&~ 1535\\
		\hline\hline
	\end{tabular}
\end{table}	
\begin{table}[htbp]
\centering
	\caption{  {\small Saturation properties used to determine the model parameters: the saturation density $n_0$, the binding energy $B_0$, the incompressibility $K_0$, symmetry energy $S_0$. 
 }  }
	\begin{tabular}{cccc}\hline\hline
	~$n_0$ [fm$^{-3}$] ~& $E_{\rm Bind}$ [MeV] ~& $K_0$ [MeV] ~& $S_0$ [MeV] ~\\
	\hline
	0.16 & 16 & 240 & 31 \\
	\hline\hline
	\end{tabular}
	\label{saturation}
\end{table}	
In addition, we use the slope parameter as an input to determine the coefficient $\lambda_{\omega\rho}$ of the $\omega$-$\rho$ mixing term. In the present analysis, we need to use the slope parameter as an input to determine the strength of the vector meson mixing (namely the parameter $\lambda_{\rho\omega}$). The estimation in Ref. \cite{universe7060182} provide the best value is $L = 57.7 \pm 19$ MeV. 

For studying this sensitivity, we first study the EOSs  
for $L = 40, 57.7, 70, 80$\,MeV with $m_0=800$ MeV fixed.

In Table.~\ref{parameter}, we summarize the values of the parameters $g_{\rho NN}$ and $\lambda_{\omega \rho}$ for several choices of the chiral invariant mass and the slope parameter. 
\begin{table}[htbp]
\centering
\caption{  {\small 
 Determined values of $\lambda_{\omega\rho}$ and $g_{\rho NN}$  
with different choices of  the chiral invariant mass $m_0$ and the slope parameter $L$.} } 
\begin{tabular}{cccccc}
\hline \hline
\multicolumn{6}{l}{$L=40$\,MeV}\\
$m_0[\mathrm{MeV}]$ & 500 & 600 & 700 & 800 & 900 \\
\hline
$\lambda_{\omega\rho}$ & 0.045 & 0.087 & 0.192 & 0.504 & 3.243 \\
$g_{\rho N N}$ & 7.31 & 7.85 & 8.13 & 8.30 & 8.43 \\
\hline
\end{tabular}
\begin{tabular}{cccccc}
\hline 
\multicolumn{6}{l}{$L=57.7$\,MeV}\\
$m_0[\mathrm{MeV}]$ & 500 & 600 & 700 & 800 & 900 \\
\hline
$\lambda_{\omega\rho}$ & 0.037 & 0.066 & 0.141 & 0.362 & 2.28 \\
$g_{\rho N N}$ & 7.31 & 7.85 & 8.13 & 8.30 & 8.43 \\
\hline 
\end{tabular}
\begin{tabular}{cccccc}
\hline 
\multicolumn{6}{l}{$L=70$\,MeV}\\
$m_0[\mathrm{MeV}]$ & 500 & 600 & 700 & 800 & 900 \\
\hline
$\lambda_{\omega\rho}$ & 0.028 & 0.045 & 0.088 & 0.211 & 1.252 \\
$g_{\rho N N}$ & 7.31 & 7.85 & 8.13 & 8.30 & 8.43 \\
\hline \hline
\end{tabular}
\begin{tabular}{cccccc}
\hline 
\multicolumn{6}{l}{$L=80$\,MeV}\\
$m_0[\mathrm{MeV}]$ & 500 & 600 & 700 & 800 & 900 \\
\hline
$\lambda_{\omega\rho}$ & 0.020 & 0.021 & 0.025 & 0.030 & 0.013 \\
$g_{\rho N N}$ & 7.31 & 7.85 & 8.13 & 8.30 & 8.43 \\
\hline \hline
\end{tabular}
\label{parameter}
\end{table}
Since the introduction of $\omega$-$\rho$ mixing does not have impacts on the normal nuclear matter construction, the coupling constants  of scalar mesons, $\bar{\mu}^2$, $\lambda_4$ and $\lambda_6$  are exactly same as those determined in Ref.~\cite{PhysRevC.103.045205},  
we only list the values for the $\lambda_{\omega\rho}, g_{\rho NN}$.

\begin{figure}[htbp]\centering
\includegraphics[width=0.9\hsize]{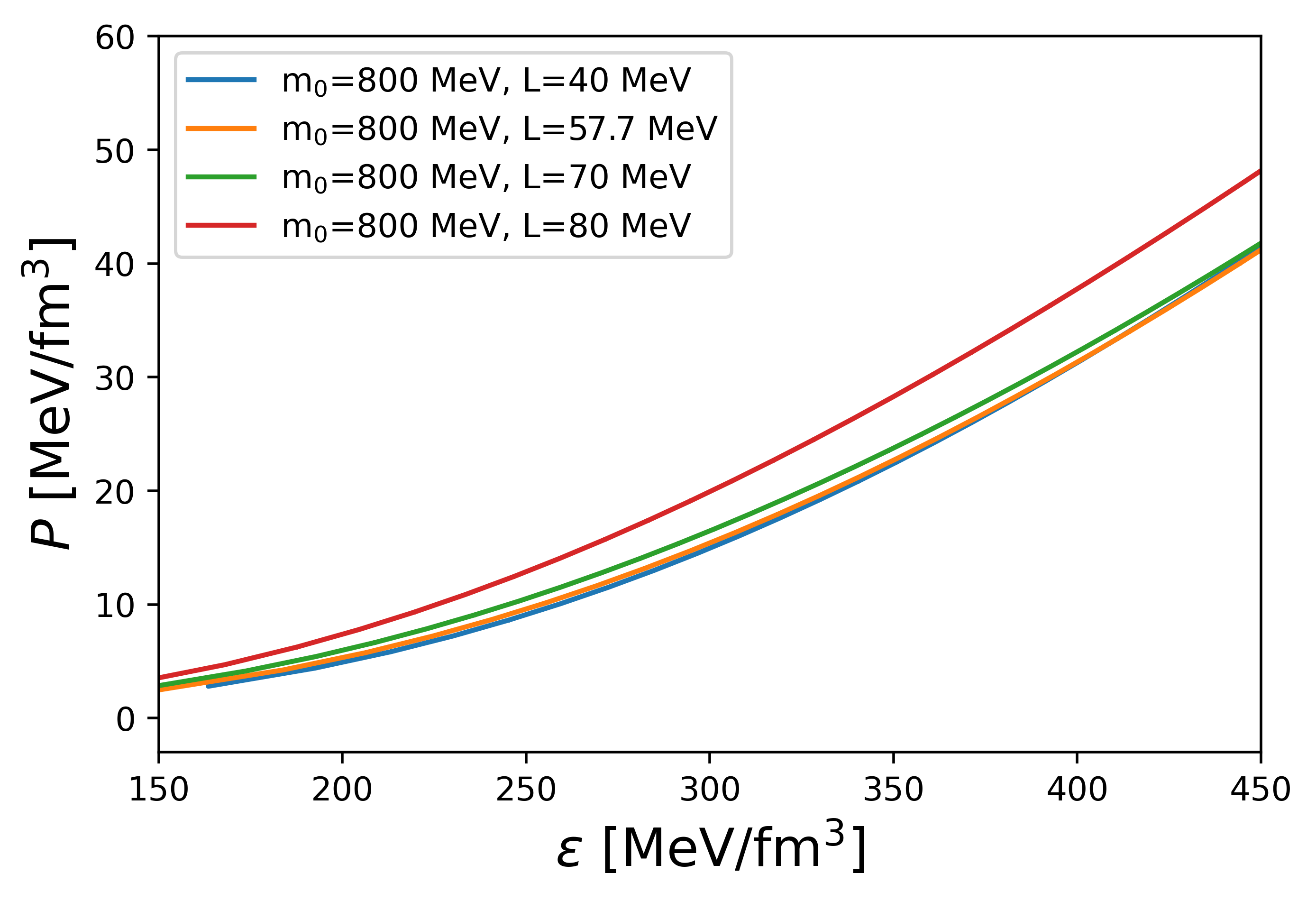}
\caption{EOS for different values of the slope parameter $L$ for $m_0 = 800$ MeV. }
\label{m_cons}
\end{figure}

The dependence on the slope parameter $L$ for $m_0=800$\,MeV is plotted in Fig.~\ref{m_cons}. 
This shows 
that the smaller $L$ leads to softer EOS as expected. As we will show later, we need vert soft EOS in the low density region to reproduce the HESS data. Then, we will take $L=40$ MeV as a typical choice in the preceding analysis. 

\begin{figure}[htbp]\centering
\includegraphics[width=0.9\hsize]{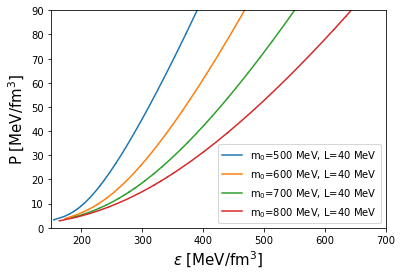}
\caption{EOS for different values of $m_0$ for $L=40$ MeV. }
\label{L_cons}
\end{figure}

We can then calculate the  EOS in the hadronic model  and the  corresponding EOS for PDM with fixing slope parameter $L=40$ MeV is shown in Fig.~\ref{L_cons}. 
From  this figure, 
we easily find that larger values of $m_0$ lead to softer EOSs.
This is understood as follows:
a greater $m_0$ leads to a weaker $\sigma$ coupling to nucleons, because a nucleon does not have to acquire its mass entirely from the $\sigma$ fields. The couplings to $\omega$ fields are also smaller because the repulsive contributions from $\omega$ fields must be balanced with attractive $\sigma$ contributions at the saturation density $n_0$. At densities larger than $n_0$, however, the $\sigma$ field reduces but the $\omega$ field increases, and these contributions are no longer balanced, affecting the stiffness of the EOS.


\subsection{QUARK MATTER EOS}
\label{NJL matter}
Following Refs.\cite{Baym:2017whm,Baym:2019iky}, we use an NJL-type quark model to describe the quark matter. 
The model includes three-flavors and U(1)$_A$ anomaly effects through the quark version of the KMT interaction. The coupling constants are chosen to be the Hatsuda-Kunihiro parameters 
which successfully reproduce the hadron phenomenology at low energy \cite{Baym:2017whm, Hatsuda:1994pi}: 
$G\Lambda^{2}=1.835, K\Lambda^{5}=9.29$ with $\Lambda=631.4\, \rm{MeV}$, see the definition below.
The couplings $g_{V}$ and $H$ characterize the strength of the vector repulsion and attractive diquark correlations whose range will be examined later 
when we discuss the NS constraints.

We can then write down the thermodynamic potential as
\begin{equation}
\begin{aligned}
\Omega_{\mathrm{CSC}}
=&\, \Omega_{s}-\Omega_{s}\left[\sigma_{f}=\sigma_{f}^{0}, d_{j}=0, \mu_{q}=0\right] \\
&+\Omega_{c}-\Omega_{c}\left[\sigma_{f}=\sigma_{f}^{0}, d_{j}=0\right],
\end{aligned}
\end{equation}
where 
the subscript 0 is attached for the vacuum values, and
\begin{align}
&\Omega_{s}=-2 \sum_{i=1}^{18} \int^{\Lambda} \frac{d^{3} \mathbf{p}}{(2 \pi)^{3}} \frac{\epsilon_{i}}{2} \label{energy eigenvalue},\\
&\Omega_{c}=\sum_{i}\left(2 G \sigma_{i}^{2}+H d_{i}^{2}\right)-4 K \sigma_{u} \sigma_{d} \sigma_{s}-g_{V} n_{q}^{2},
\end{align}
with $\sigma_{f}$ being the chiral condensates, $d_{j}$ denotes for diquark condensates, and $n_{q}$ denotes for the quark density. 
In Eq.(\ref{energy eigenvalue}), $\epsilon_{i}$ are energy eigenvalues obtained from inverse propagator in Nambu-Gorkov bases
\begin{equation}
S^{-1}(k)=\left(\begin{array}{lc}
\gamma_{\mu} k^{\mu}-\hat{M}+\gamma^{0} \hat{\mu} & \gamma_{5} \sum_{i} \Delta_{i} R_{i} \\
-\gamma_{5} \sum_{i} \Delta_{i}^{*} R_{i} & \gamma_{\mu} k^{\mu}-\hat{M}-\gamma^{0} \hat{\mu}
\end{array}\right),
\end{equation}
where
\begin{equation}
\begin{aligned}
&M_{i} =m_{i}-4 G \sigma_{i}+K\left|\epsilon_{i j k}\right| \sigma_{j} \sigma_{k}, \\
&\Delta_{i} =-2 H d_{i} ,\\
&\hat{\mu} =\mu_{q}-2 g_{V} n_{q}+\mu_{3} \lambda_{3}+\mu_{8} \lambda_{8}+\mu_{Q} Q,\\
&(R_{1}, R_{2}, R_{3})=(\tau_{7}\lambda_{7}, \tau_{5}\lambda_{5}, \tau_{2}\lambda_{2}).
\end{aligned}
\end{equation}
$S^{-1}(k)$ is $72\times72$ matrix in terms of the color,
flavor, spin, and Nambu-Gorkov basis, which has 72 eigenvalues. $M_{u,d,s}$ are the constituent masses of $u, d, s$ quarks and $\Delta_{1,2,3}$ are the gap energies. 
The $\mu_{3,8}$ are the color chemical potentials which will be tuned to achieve the color neutrality. 
The total thermodynamic potential including the effect of leptons is 
\begin{equation}
\Omega_{\mathrm{Q}}=\Omega_{\mathrm{CSC}}+\sum_{l=e, \mu} \Omega_{l}.
\end{equation}
The mean fields are determined from the gap equations,
\begin{equation}
0=\frac{\partial \Omega_{\mathrm{Q}}}{\partial \sigma_{i}}=\frac{\partial \Omega_{\mathrm{Q}}}{\partial d_{i}},
\end{equation}
From the conditions for electromagnetic charge neutrality and color charge neutrality, we have
\begin{equation}
n_{j}=-\frac{\partial \Omega_{\mathrm{Q}}}{\partial \mu_{j}}=0,
\end{equation}
where $j = 3,8, Q$. 
The baryon number density $n_{B}$ is determined as
\begin{equation}
n_{q}=-\frac{\partial \Omega_{\mathrm{Q}}}{\partial \mu_{q}},
\end{equation}
where $\mu_{q}$ is $1/3$ of the baryon number chemical potential. After determined all the values, we obtain the pressure as
\begin{equation}
P_{\mathrm{Q}}=-\Omega_{\mathrm{Q}}.
\end{equation}
%


\section{STUDY OF PROPERTIES OF NS}
\label{sec-NS}

In this section,  following Ref.~\cite{PhysRevC.103.045205} 
we construct a unified EOS by connecting the EOS obtained in the PDM introduced in Sec.~\ref{sec:PDM matter} 
and the EOS of NJL-type quark model given in Sec.~\ref{NJL matter}, and solve the TOV equation~\cite{Tolman:1939jz,Oppenheimer:1939ne} 
to obtain the NS mass-radius  ($M$-$R$)  relation. 
As for the interplay between nuclear and quark matter EOS, see, e.g., Ref.~\cite{Kojo:2020krb} for a quick review that classifies types of the interplay.


\subsection{Construction of unified EOS}

%
\begin{table}[tbh]
\begin{center}
\begin{tabular}{c|c|c|c}
\hline
\hline
$0\leq n_{B}<0.5 n_{0} $ & $0.5 n_{0}\leq n_{B}\leq 2n_{0}$ & $2n_{0}<n_{B}<5n_{0}$ & $n_{B}\geq 5n_{0}$\\
\hline
\rm{Crust} & \rm{PDM} & \rm{Interpolation} & \rm{NJL}\\
\hline
\hline
\end{tabular}
\end{center}
\caption{Unified EOS composed of four part.}
\label{UniEOS}
\end{table}
In our unified equations of state as in Table.\ref{UniEOS},
we use the BPS (Baym-Pethick-Sutherland) EOS \cite{Baym:1971pw} as a crust EOS for $n_B \lesssim 0.5n_0$. 
From $n_B \simeq 0.5n_0$ to $2n_0$ we use our PDM model to describe a nuclear matter.  
We limit the use of our PDM up to $2n_{0}$ so that baryons other than ground state nucleons, such as the negative parity nucleons or hyperons, do not show up in matter. Beyond $2n_0$ nuclear regime, we assume a crossover from the nuclear matter to quark matter, 
and use a smooth interpolation to construct the unified EOS. 
We expand the pressure as a fifth order polynomial of $\mu_{B}$ as
\begin{equation}
P_{\mathrm{I}}\left(\mu_{B}\right)=\sum_{i=0}^{5} C_{i} \mu_{B}^{i},
\end{equation}
where $C_{i}$  ($i=0,\cdots, 5$) are parameters  to be determined from boundary conditions  given by 
\begin{equation}
\begin{aligned}
&\left.\frac{\mathrm{d}^{n} P_{\mathrm{I}}}{\left(\mathrm{d} \mu_{B}\right)^{n}}\right|_{\mu_{B L}}=\left.\frac{\mathrm{d}^{n} P_{\mathrm{H}}}{\left(\mathrm{d} \mu_{B}\right)^{n}}\right|_{\mu_{B L}}, \\
&\left.\frac{\mathrm{d}^{n} P_{\mathrm{I}}}{\left(\mathrm{d} \mu_{B}\right)^{n}}\right|_{\mu_{B U}}=\left.\frac{\mathrm{d}^{n} P_{\mathrm{Q}}}{\left(\mathrm{d} \mu_{B}\right)^{n}}\right|_{\mu_{B U}}, \quad(n=0,1,2),
\end{aligned}
\end{equation}
with $\mu_{BL}$  being 
the chemical potential corresponding to $n_{B}=2n_{0}$ and $\mu_{BU}$ to $n_{B}=5n_{0}$. We demand the matching up to the second order derivatives of pressure at each boundary.
The resultant interpolated EOS must satisfy the thermodynamic stability condition,
\begin{align}
\chi_B = \frac{\, \partial^2 P \,}{\, (\partial \mu_B )^2 \,} \ge 0 \,,
\end{align}
and the causality condition,
\begin{equation}
c_{s}^{2}=\frac{\, \mathrm{d} P \,}{\mathrm{d} \varepsilon} 
= \frac{n_{B}}{\mu_{B} \chi_{B}} \le 1 \,,
\end{equation}
which means that the sound velocity is smaller than the light velocity.
These conditions restrict the range of quark model parameters $(g_V, H)$ for a given nuclear EOS and a choice of $(n_L, n_U)$. We exclude interpolated EOSs which do not satisfy the above-mentioned constraints.

\subsection{Mass-Radius relation}

In this section, we calculate mass-radius relation of NSs by using the unified EOS constructed in the previous section for  the  PDM with different parameter choices of chiral invariant mass $m_0$ and slope parameter $L$.  

\begin{figure*}[t]\centering
\begin{subfigure}{0.47\hsize}\centering
	\includegraphics[width=0.95\hsize]{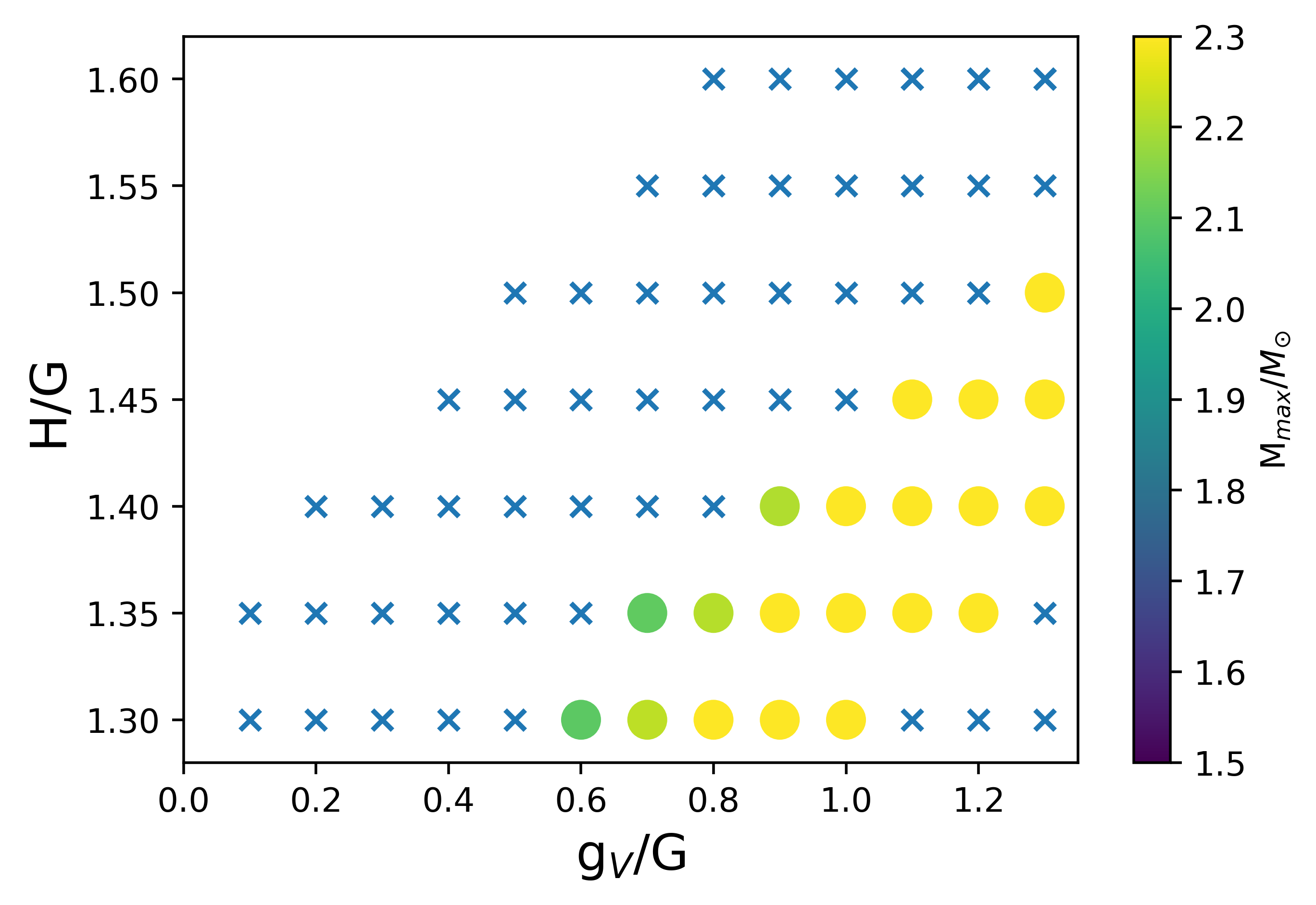}
	\caption{$m_0$=500 MeV}
	\label{fig-baryon-psi}
\end{subfigure}
\begin{subfigure}{0.47\hsize}\centering
	\includegraphics[width=0.95\hsize]{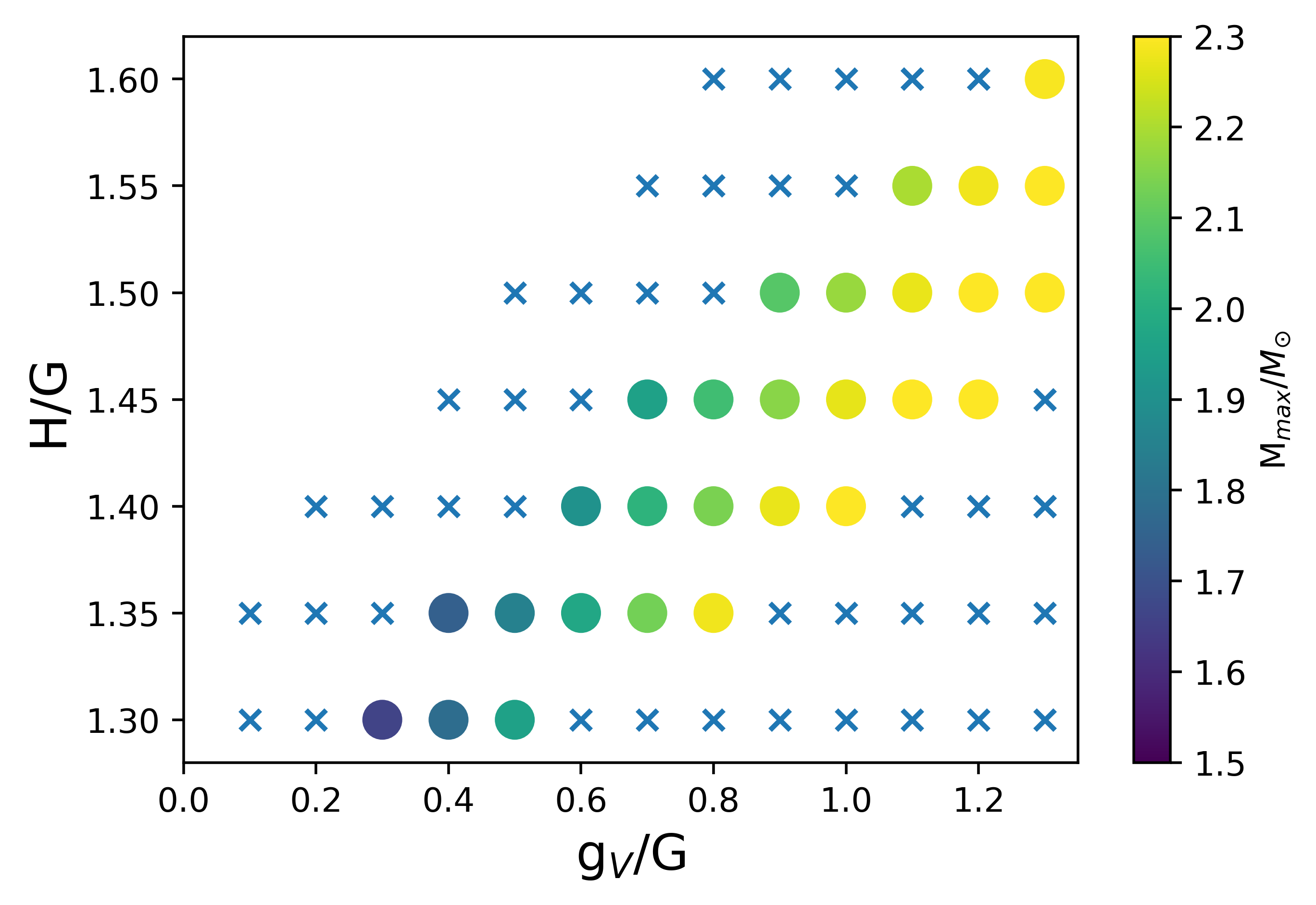}
	\caption{$m_0$=600 MeV}
	\label{fig-baryon-psi}
\end{subfigure}
\begin{subfigure}{0.47\hsize}\centering
	\includegraphics[width=0.95\hsize]{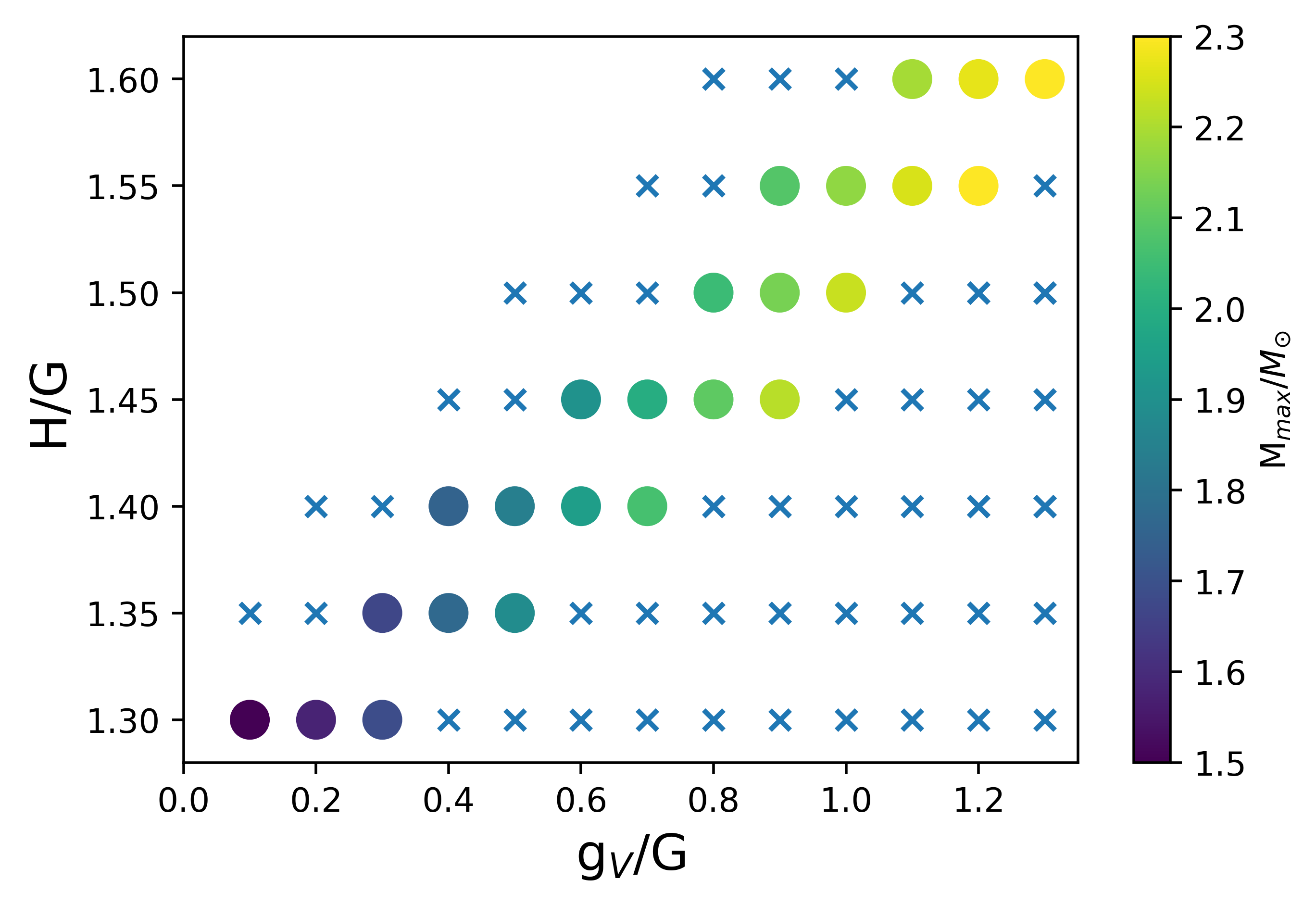}
	\caption{$m_0$=700 MeV}
	\label{fig-baryon-psi}
\end{subfigure}
\begin{subfigure}{0.47\hsize}\centering
	\includegraphics[width=0.95\hsize]{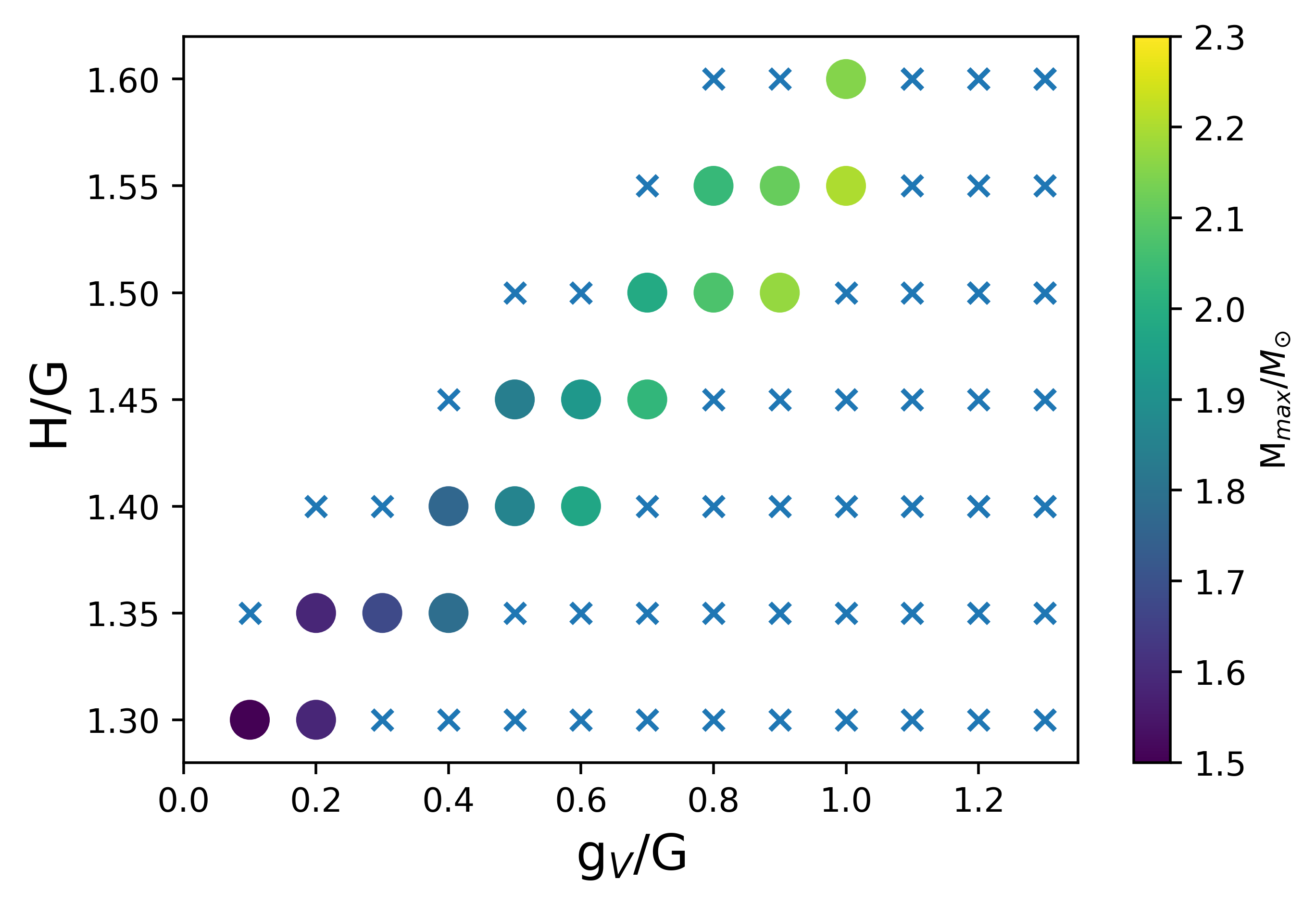}
	\caption{$m_0$=800 MeV}
	\label{fig-baryon-psi}
\end{subfigure}
\begin{subfigure}{0.47\hsize}\centering
	\includegraphics[width=0.95\hsize]{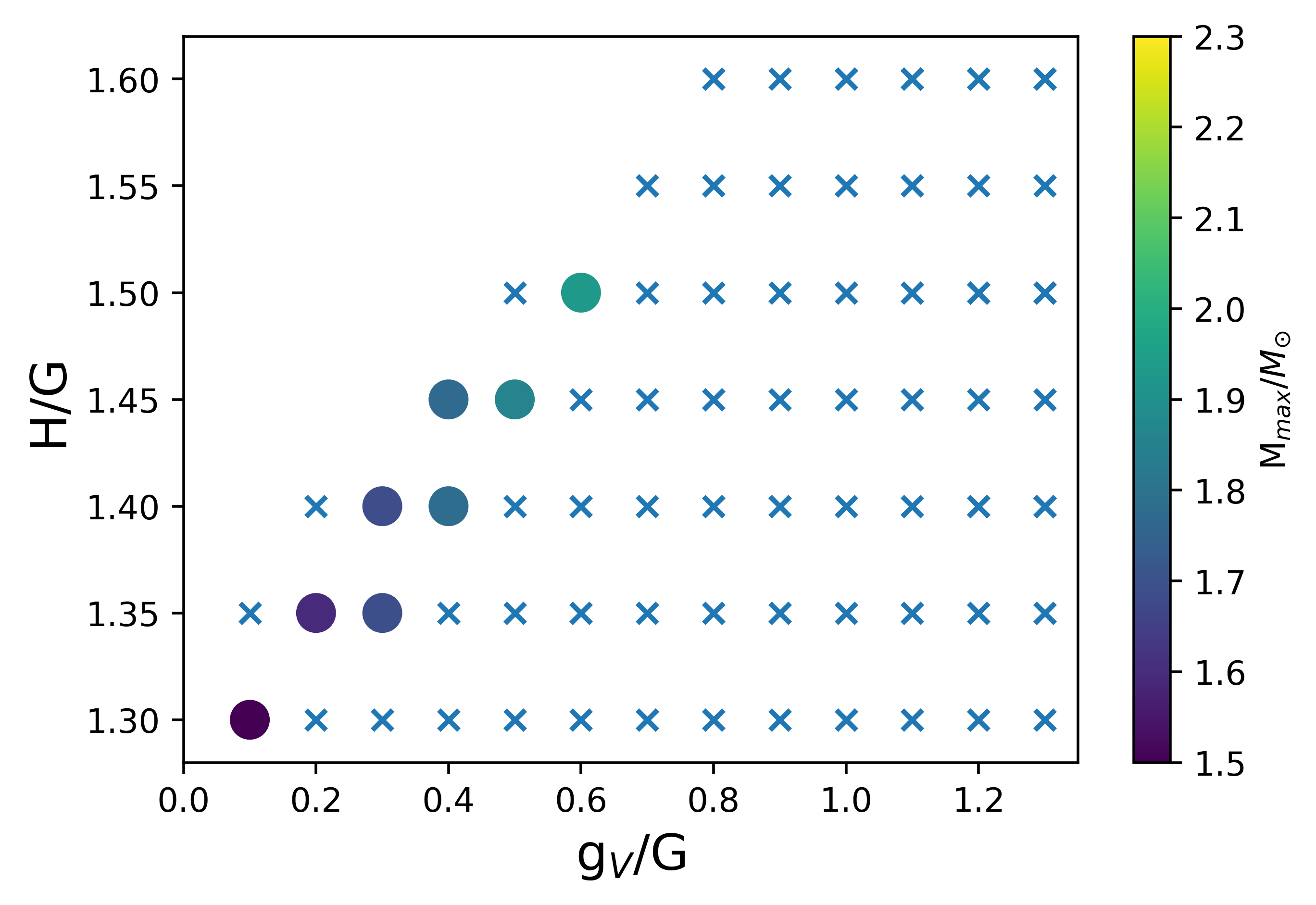}
	\caption{$m_0$=900 MeV}
	\end{subfigure}
\caption[]{
Allowed combination of ($H, g_V$) values for $m_0=500, 600, 700, 800, 900$ MeV when $L=40$ MeV. Cross mark indicates that the combination of ($H, g_V$) is excluded by the causality constraints. Circle indicates that the combination is allowed. The color shows the maximum mass of NS obtained from the corresponding parameters, as indicated by a vertical bar at the right side of each figure.}
\label{H_gv}
\end{figure*}

First, we study whether the smooth connection is realized depending on the parameters $H$ and $g_V$ in the NJL-type quark model as shown in Fig. \ref{H_gv} for PDM with $L=40$\,MeV. 
For each combination of $(H, g_V)$, 
the cross mark are the parameter choices forbidden by the causality  and thermodynamic stability conditions.
For possible choices of $(H,g_V)$, we determine the maximum mass of a NS, which is indicated by the color in Fig.~\ref{H_gv}. This shows that a larger $g_V$ or/and a smaller $H$ leads to a larger maximum mass. For $m_0=900$ MeV, the maximum mass for all the choices of $(H,g_{V})$ are below 2$M_{\odot}$, leading to the conclusion that $m_0=900$ should be excluded when slope parameter is chosen to be $L=40$ MeV.

\begin{figure}[htbp]
\includegraphics[width=\hsize]{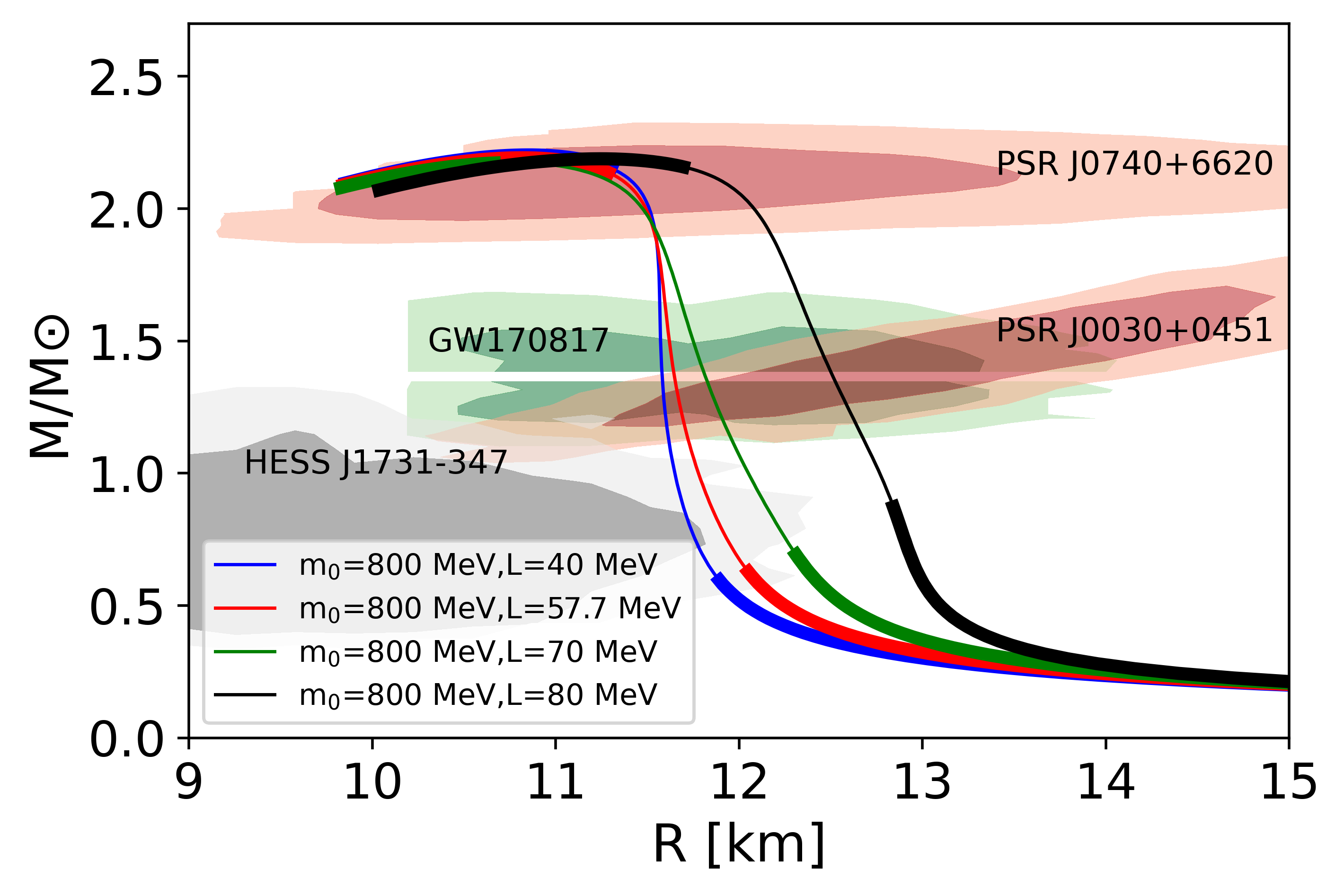}
\caption{Mass-radius relations for same $m_0=800$ MeV in different PDM sets. Black curve is connected to the NJL parameters (H, $g_V$)/G = (1.5, 1); green curve to (H, $g_V$)/G = (1.55, 1); red curve to (H, $g_V$)/G = (1.55, 1); blue curve to (H, $g_V$)/G = (1.55, 1).}
\label{mr_changeL}
\end{figure}
In Fig.~\ref{mr_changeL}, we fix the value of $m_0$ with different choice of $L$ and calculate the corresponding mass-radius curves, 
 where the values of 
($H, g_V$) are chosen to have the stiffest EOS.  In this figure, 
the thick part indicates that the density region is smaller than $2n_0$ or larger than 5$n_0$ and the thin line indicates the interpolated region. From the figure, for $m_0=800$ MeV, the radius for  $L=40$ MeV, $M\simeq 1.4M_{\odot}$ is about $11.5$\,km while the result of $L=80$\,MeV about $12.6$\,km. This result  indicates that EOSs are softened by the effect of the $\omega\rho$ interaction. One can see that the $M$-$R$ curve for $L=40$\,MeV satisfies the constraint from the HESS J1731-347 observation. We note that $L=40$\,MeV is consistent with the one obtained in Ref.~\cite{universe7060182}, 
due to a large ambiguity. Precise determination of slope parameter in
future will help us to 
further constrain the NS properties. 

\begin{figure}[htbp]\centering
\includegraphics[width=\hsize]{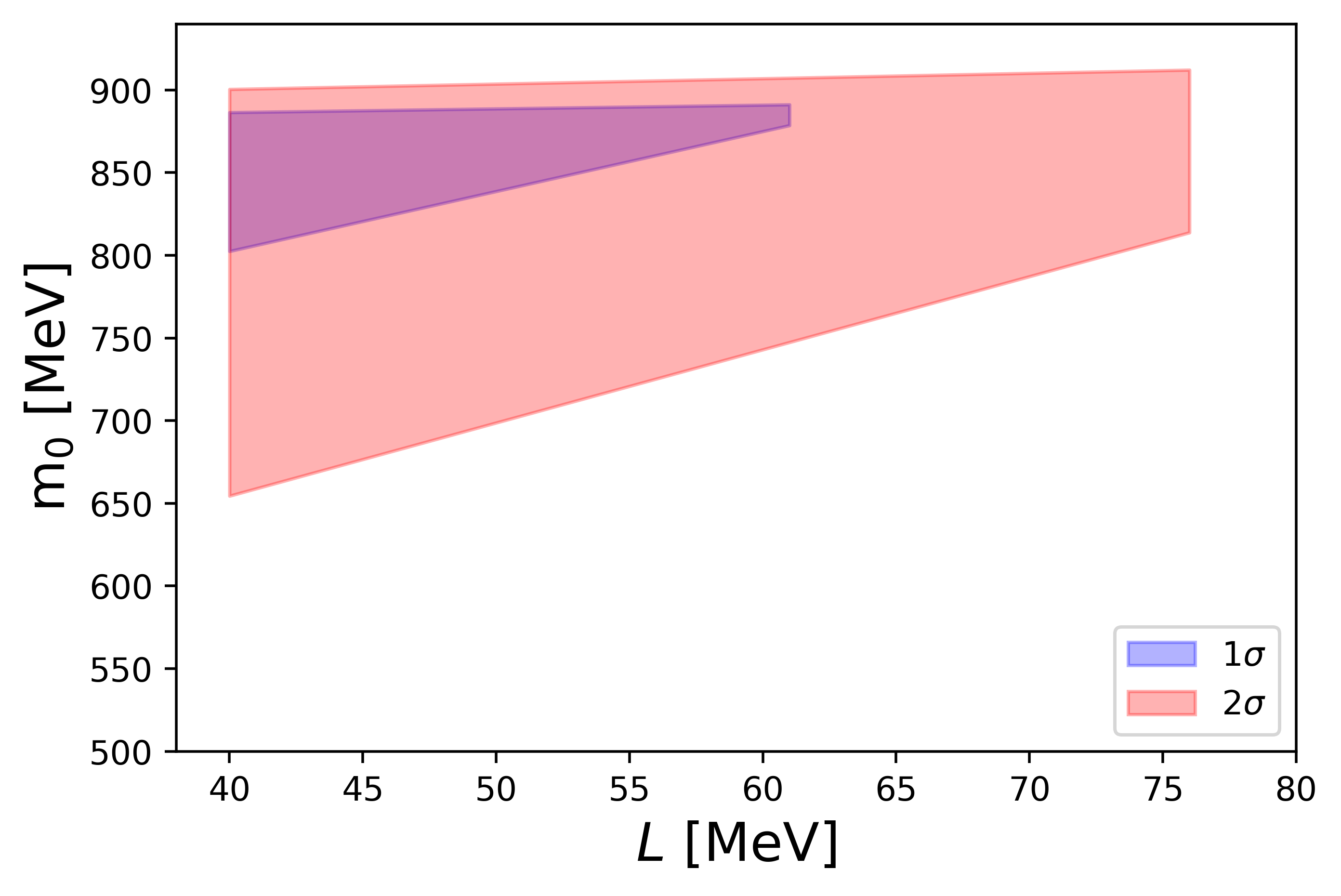}
\caption{Allowed region for $m_0$ and $L$. Within the shadowed region, the M-R curve satisfy all the constraint from the NS observation within the error of $1\sigma$ or $2\sigma$. 
}
\label{m0_L}
\end{figure}

To achieve a NS with small radius, the outer core EOS (Density around $1n_0$-$2 n_0$) is extremely important, since it directly connects to the radius of a neutron star. In our model, the chiral invariant mass $m_0$ and the slope parameter $L$ are two factors which have impacts on the outer core EOS.
We then treat them as free parameters and compare the corresponding $M$-$R$ curves with NS constraints from NICER, gravitational wave detection and HESS. We show the allowed region of $m_0$ and $L$ satisfying all the 
observational  constraints in $1\sigma$ and $2\sigma$ range as in Fig. \ref{m0_L}. 
Under this parameter space favoring large $m_0$ and small $L$, HESS J1731-347 can be considered as the lightest NS. 
%


\section{SUMMARY AND DISCUSSIONS}
\label{summary}

In this study, we use parity doublet model together with NJL-type model within the framework of relativistic mean-field model to describe low-mass neutron stars. We construct EOS for NS matter by interpolating the EOS obtained in the PDM and the one in the NJL-type model 
with  assuming the crossover from hadronic matter to quark matter.  In the calculation of the NS mass-radius relation, we find outer core EOS is crucial to determine the radius of a NS. Consequently, the choices of chiral invariant mass $m_0$ and slope parameter $L$  which 
describe the properties of the uniform nuclear matter  are 
essential. We treat $m_0$ and $L$ as two free parameters and find the parameter space enable us to explain the HESS J1731-347 as a neutron star as in Fig. \ref{m0_L}.

We note here that 
the typical estimate of $L$ falls  
within the range of $40$-$80$ MeV, as indicated by various studies\cite{universe7060182, Drischler:2020hwi, Tews:2016jhi}. However, there are also other estimates such as $L=(109 \pm 36.41)$ MeV derived from the analyses of neutron skin thickness from PREX-2 experiment. There is still large ambiguities about the value of slope parameter.
In the present research,  we follow Ref.~\cite{universe7060182} 
as the baseline to set $L=57.7 \pm 19$ MeV and 
study the corresponding mass-radius relation. If future experiment show the value of slope parameter is large, we can come to the conclusion that HESS J1731-347 cannot be explained as a NS within the present model.

\begin{figure}[htbp]\centering
\includegraphics[width=\hsize]{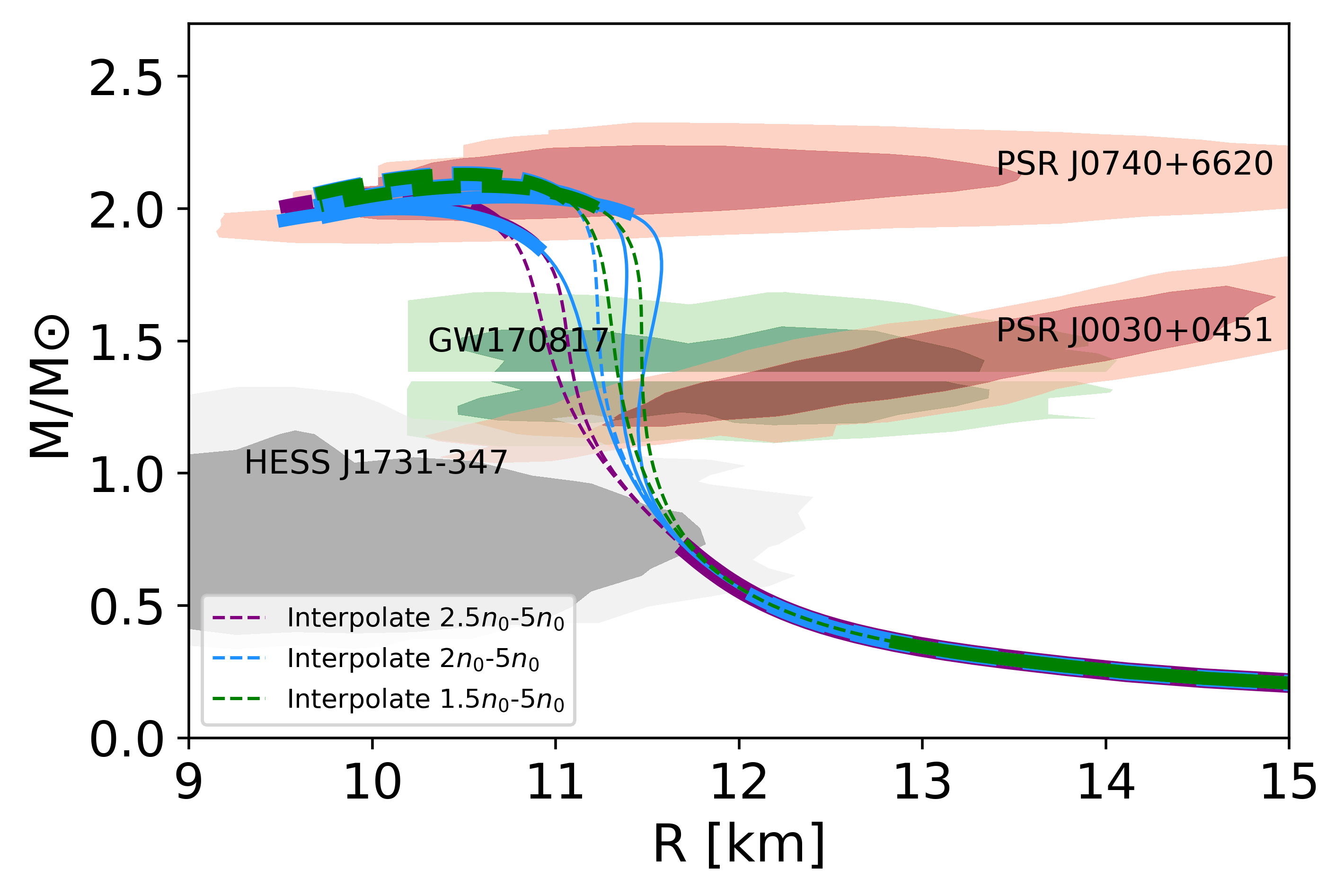}
\caption{Mass-radius relations for $m_0 = 850$\,MeV, $L = 40$\,MeV and corresponding  curves for central density. 
Different colors indicate 
different interpolation range.
}
\label{mr_850}
\end{figure}

As studied in Refs.\cite{Masuda:2012kf, Masuda:2012ed,Baym_2018}, the validity of pure hadronic descriptions at $n_B \geq 2n_0$ are questionable as nuclear many-body forces are very important, implying that quark descriptions are required even before the quark matter formation. In this study, we choose the interpolation point to be $2n_0$ and the ambuguity from the interpolation point is disscussed in Fig.~\ref{mr_850}. In this figure, we show the $M$-$R$ curves for $m_0=850\,$MeV and $L=40\,$MeV with  
changing the interpolation range from $2n_0$-$5 n_0$ to $1.5 n_0$-$5n_0$ and $2.5n_0$-$5n_0$.
We can easily  see
that the ambiguity from the interpolation point is very limited: at the mass about $1M_{\odot}$, the radius shifts are only about $0.1$ 
km.

\begin{figure}[htbp]
\includegraphics[width=\hsize]{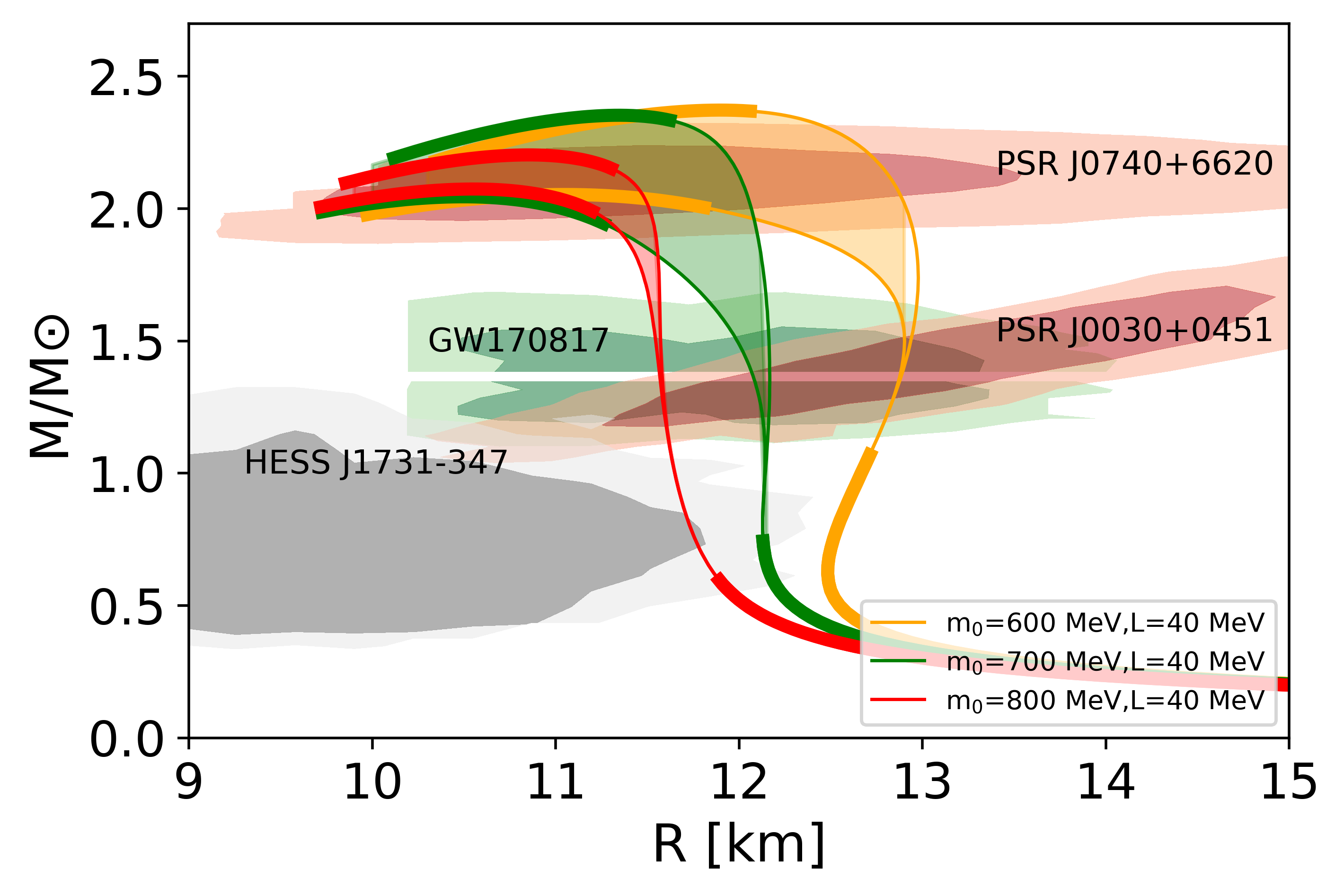}
\caption{
Mass-Radius relations for $m_0 = 600$, $700$, $800$\,MeV with $L=40\,$MeV.
Orange curves are for $(H, g_V)/G = (1.55, 1.3)$ and $(1.45, 0.8)$; green curves for $(H, g_V)/G = (1.6, 1.3)$ and $(1.5, 0.8)$; red curves for $(H, g_V)/G = (1.55, 1)$, $(1.5, 0.8)$.
}
\label{mr_changem0}
\end{figure}

\begin{figure}[htbp]\centering
\includegraphics[width=\hsize]{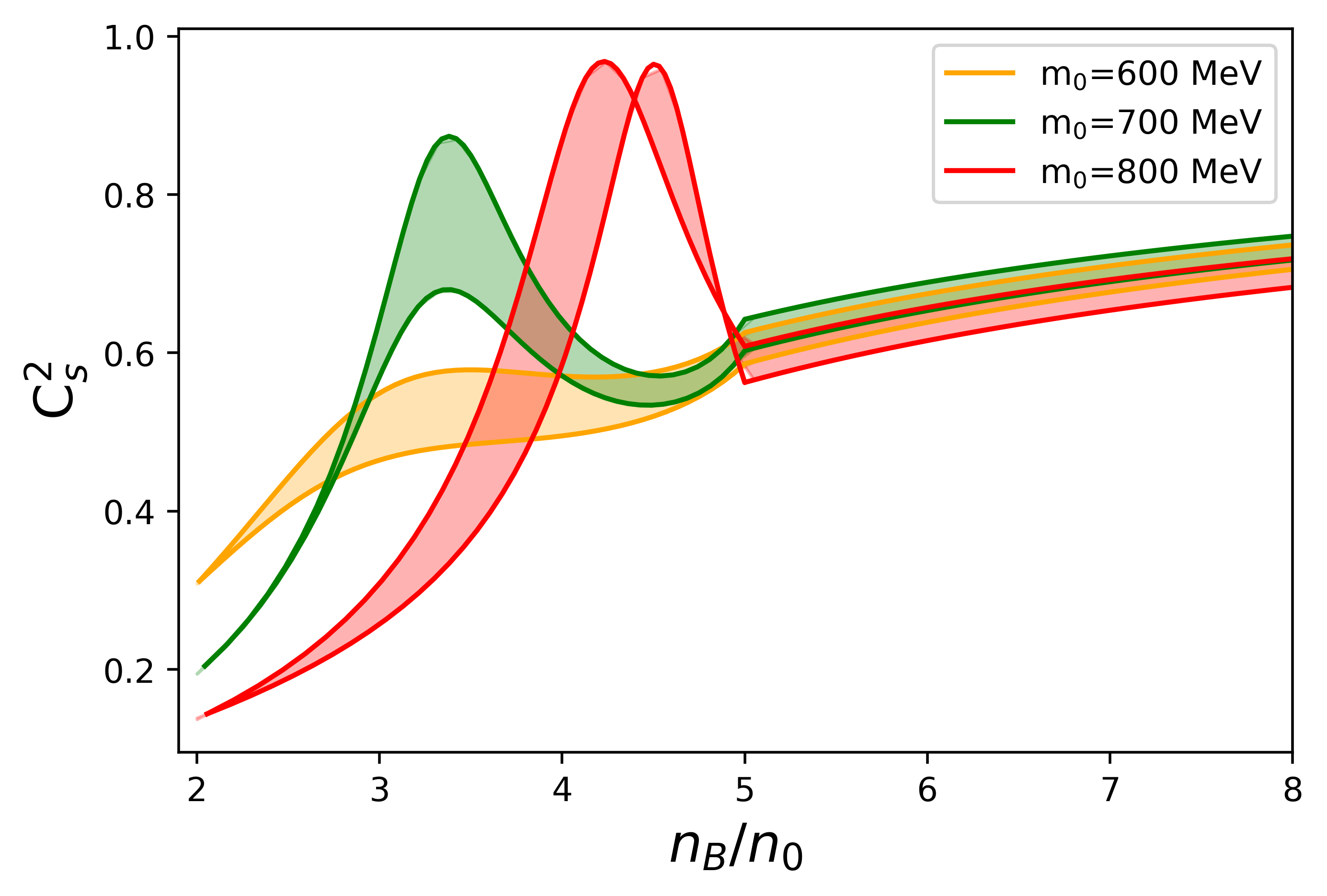}
\caption{
Sound velocity for  $m_0=600$, $700$ and $800$ \,MeV. The NJL parameters are the same as Fig. \ref{mr_changem0}.  
}
\label{mr_changem0_sv}
\end{figure}

In Fig.~\ref{mr_changem0}, we fix the value of slope parameter as $L=40$\,MeV 
and vary the value of $m_0$ as  $m_0=600, 700, 800$ MeV.
We
choose the values of 
$(H, g_V)$ parameters to produce the most stiff and the most soft EOSs 
satisfying $2M_{\odot}$ constraint. For $m_0 = 700, 800$ MeV, the rather soft hadronic EOSs are connected with rather stiff quark EOSs satisfying $2M_{\odot}$ constraint, resulting 
a peak of the density dependence of  
sound velocity, as shown in Fig.~\ref{mr_changem0_sv}.
However, for $m_0 = 600$ MeV, the rather stiff hadronic EOS is used to connect with stiff quark EOSs, resulting just a bump-like structure.
Besides, we find that
the onset density of the sound velocity peak is larger for larger $m_0$. 
Reference~\cite{Marczenko:2022jhl} 
pointed out that the appearance of the maximum in the speed of sound in the interior of NSs might indicate the change of medium composition, from hadronic to quark or quarkyonic matter. 
They estimate the critical density where baryons begin to overlap as $n_{c}^{per}=1.22/V_0, V_0 = (4/3)\pi R^{3}_0$\cite{Braun-Munzinger:2014lba}. After using experimental value of the proton radius $R_0 = 0.9 \pm 0.05$ fm\cite{Dey:2014tfa,PhysRevLett.95.182001},  the critical density is calculated as $n_{c}^{per} = 0.57^{+0.12}_{-0.09}$ fm$^{-3}$. 
When we require that the peak density of the sound velocity in the present analysis should satisfy $0.48 \leq n_B^{\rm peak} \leq 0.69$, i.e. $3 \leq n_B^{\rm peak}/n_0 \leq 4.3$, we obtain the constraint to the chiral invariant mass as $600 \lesssim m_0 \lesssim 800$\,MeV for $L=40$ MeV.

\section*{Acknowledgement}
The work of  B.G., and M.H. are supported in part by JSPS KAKENHI Grant Nos.~20K03927, 23H05439 and 24K07045.
B.G. is also supported by JST SPRING, Grant No. JPMJSP2125.  B.G. would like to take this opportunity to thank the
“Interdisciplinary Frontier Next-Generation Researcher Program of the Tokai
Higher Education and Research System.”

\bibliography{ref_3fPDM_2022.bib}

\end{document}